\documentclass{iopart}

\usepackage{graphicx} 
\usepackage[english]{babel}
\usepackage[pstricks,mediumqspace,squaren]{SIunits} 

\newcommand{\thold}{t_{\textrm{hold}}}
\newcommand{\Nbec}{N_{\textrm{BEC}}}
\newcommand{\abg}{a_{\textrm{bg}}}
\newcommand{\aBohr}{a_{\textrm{Bohr}}}
\newcommand{\aevap}{a_{\textrm{evap}}}
\newcommand{\ai}{a_{\textrm{i}}}
\newcommand{\af}{a_{\textrm{f}}}
\newcommand{\acrit}{a_{\textrm{crit}}}

\begin{document}
\title[special issue on cold and ultracold molecules; %
  editors: Jun Ye and Lincoln Carr]{%
  {Coherent collapses of dipolar Bose-Einstein condensates for different
    trap geometries}}

\newcommand{\UniStuttgartAddress}{%
  5.~Physikalisches Institut, %
  Universit{\"a}t Stuttgart, %
  Pfaffenwaldring 57, %
  70569 Stuttgart, %
  Germany}

\newcommand{\SaitoAddress}{%
  Department of Applied Physics and Chemistry, %
  The University of Electro-Communications, %
  Tokyo 182-8585, %
  Japan}

\newcommand{\YukiAddress}{%
  Department of Physics, %
  University of Tokyo, %
  Tokyo 113-033, %
  Japan}

\newcommand{\UedaAddress}{%
  ERATO Macroscopic Quantum Project, %
  JST, %
  Tokyo 113-8656, %
  Japan}

\author{%
  J Metz$^1$, %
  T Lahaye$^1$, %
  B Fr{\"o}hlich$^{1}$, %
  A Griesmaier$^1$, %
  T Pfau$^1$, %
  H Saito$^2$, %
  Y Kawaguchi$^3$, %
  and %
  M Ueda$^{3,4}$ %
} %
\address{$^1$\UniStuttgartAddress} %
\address{$^2$\SaitoAddress} %
\address{$^3$\YukiAddress} %
\address{$^4$\UedaAddress} %

\ead{t.pfau@physik.uni-stuttgart.de}


\date{\today}

\begin{abstract}
  We experimentally investigate the collapse dynamics of dipolar
  Bose-Einstein condensates of chromium atoms in different harmonic
  trap geometries, from prolate to oblate. The evolutions of the
  condensates in the unstable regime are compared to three-dimensional
  simulations of the Gross-Pitaevskii equation including three-body
  losses.  In order to probe the phase coherence of collapsed
  condensates, we induce the collapse in several condensates
  simultaneously and let them interfere.
\end{abstract}


\section{Introduction}
A collapse is a fast, collective phenomenon consisting in the
destruction of a multi-particle system happening abruptly on the time
scale which governs the ``usual dynamics''. One example is the
gravitational core collapse initiating a supernova. Happening within
milliseconds, its duration is negligible compared to any of the time
scales related to the preceding fusion stages~\cite{Woosley:02}.

In contrary to a supernova, where the experimenter is condemned to be
an observer only, Bose-Einstein condensates (BECs) are excellent
adjustable systems.  Tailoring both the external confining potential
and the interaction between the atoms allows to control the properties
of the condensate. It is not only an ideal system to study questions
from condensed matter
physics~\cite{Fattori:08Osci,Levy:07,Greiner:02,Billy:08,Paredes:04,Kinoshita:04,Hadzibabic:06,Bloch:08},
but the dynamics of a collapse as well.

Collapsing condensates were first observed in $^7$Li~\cite{Gerton:00}
and $^{85}$Rb~\cite{Donley:01}. Both systems are characterized by the
contact interaction, which is described by a single parameter, the
$s$-wave scattering length $a$. From a simple model~\cite{Dalfovo:99}
minimizing the Gross-Pitaevskii energy functional using a gaussian
ansatz for the wave function, one can understand the instability
threshold. The energy functional consists of three terms: (\emph{i})
the kinetic energy, which, in the sub-micro Kelvin range of almost
pure condensates, is equal to the quantum pressure (arising from the
Heisenberg uncertainty principle in the trap potential), (\emph{ii})
the harmonic trapping potential, and (\emph{iii}) the interaction
energy. For repulsive contact interaction ($a>0$) a global minimum
exists, and the condensate is stable. For small enough attractive
interactions ($a<0$), only a \emph{local} minimum exists at finite
size; the condensate is metastable. Finally, for sufficiently
attractive interaction the local minimum vanishes. The potential
energy of the BEC can be lowered without bound by contraction.  Thus,
the condensate is unstable.

As the density increases, this model is too simple to describe the
physics of the unstable cloud -- atom losses due to three-body
collisions have to be included.  Being negligible at low densities,
the three-body collision rate rapidly increases as the cloud
shrinks. A three-body collision allows for the production of a dimer,
where the third contributing atom is needed to fulfill energy and
momentum conservation.  The binding energy which is absorbed by the
atom and the molecule in form of kinetic energy is sufficient for both
to escape from the trap. Hence, instead of reaching a fully contracted
``point-like'' state, more and more atoms are lost, so that eventually
the quantum pressure dominates over the remaining interaction
energy. The dynamics inverts, the atoms accelerate outwards to the
\emph{new} equilibrium state. Because the three-body losses have
changed the total energy, this new equilibrium state differs from the
initial one.

Inducing the collapse in a condensate with non-negligible dipolar
interactions changes the dynamics completely.  While the contact
interaction is short-range and spherically symmetric, the dipolar
interaction is long-range and cylindrically symmetric (partially
attractive and partially repulsive).  Therefore, dipolar condensates
exhibit many novel phenomena even in non-collapsing
systems~\cite{Baranov:02,Goral:02PRL,Baranov:08Dipole, 
  Santos:03, 
  ODell:04,
  Ronen:07,Dutta:07, 
  Yi:06StructuredVortex, 
  Cooper:05,Zhang:05, 
  Kawaguchi:06,Santos:06,Yi:06}. 

The anisotropy of the dipolar interaction can also be used to change
the mechanism for the collapse -- either driving or stabilizing the
collapse: In a cigar-shaped trap with the dipoles oriented along the
long axis the atoms experience predominantly the attractive part of
the dipolar mean-field potential~\cite{Koch:08}. Hence, the collapse
is \emph{driven} by the dipolar interaction, while the contact
interaction stabilizes the condensate. The instability occurs at
positive scattering length. This situation is reversed if the dipoles
are confined in a pancake-shaped trap. Now their dipolar interaction
is essentially repulsive. The dipolar interaction \emph{stabilizes}
the condensate and the condensate is able to withstand negative
scattering lengths for which a purely contact interacting BEC would
have become unstable already~\cite{Koch:08}. The instability occurs at
negative scattering length.  This simple consideration strongly
suggests that different trap geometries result in different collapse
dynamics.

In this paper we study the collapse of a chromium condensate under the
influence of magnetic dipole
interaction~\cite{Stuhler:05,Lahaye:07,Koch:08}. After describing our
experimental procedure to induce the collapse, we compare the
experimental data for different trapping potentials with 3D
simulations of the Gross-Pitaevskii equation (GPE) including
three-body losses. We show that this model provides a good description
of the experimental data. This is not clear \emph{a priori} as one
might expect that the collapse dynamics induces many-body quantum
correlations, while the GPE is a mean-field description, not taking
correlations into account. Finally, we prove that the collapsed atomic
cloud contains a remnant \emph{condensate} by probing its phase
coherence.  Although simulations for purely contact interacting BECs
have shown~\cite{Saito:02} that the observed bursts and jets of a
collapsing condensate should be coherent, this had not been tested
experimentally to date.

\section{Experimental sequence to produce collapsing dipolar condensates}
\begin{figure}[t]
  \centering
  \includegraphics[scale=1]{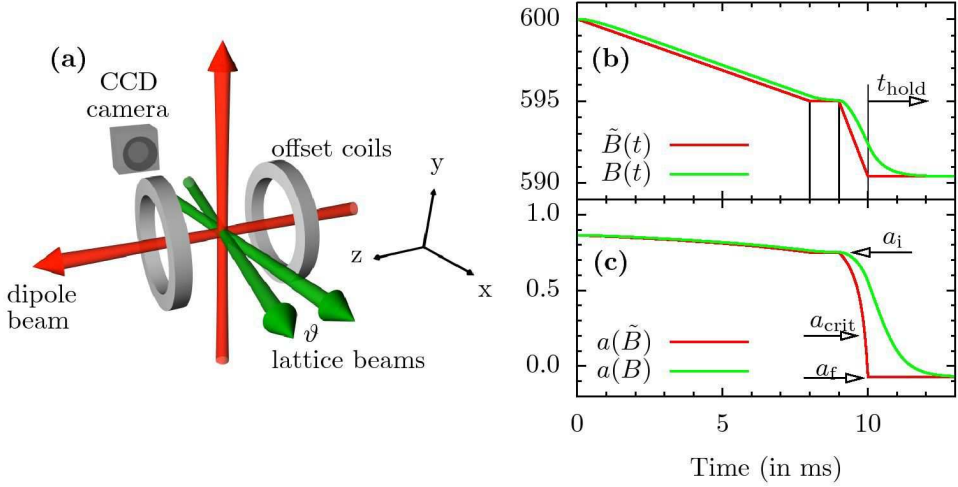}
  \caption{(a) Experimental setup: We realize different confining
    potentials by superimposing a one-dimensional optical lattice
    (green) and a crossed dipole trap (red). The offset coils produce
    the magnetic field for the Feshbach resonance. %
    (b) Eddy currents retard the magnetic field $B$ at the position of
    the atoms (green) with respect to the ramps of the offset coils
    (red). The time $\thold$ is defined as the additional holding time
    of the atoms in the optical trap after finishing the second
    magnetic field ramp, before the time-of-flight. %
    (c) The corresponding scattering length in units of $\abg$. }
  \label{fig:Exp_Setup}
  \label{fig:EddyCurrent}
\end{figure}

In order to create a condensate dominated by the dipolar interaction,
we exploit the broadest of the observed Feshbach resonances in
$^{52}$Cr~\cite{Werner:05}. In the vicinity of this Feshbach
resonance, the scattering length varies with the applied magnetic
field $B$ as~\cite{Koehler:06} %
\begin{equation} %
  \label{eq:FB_scattering_length_scaling} %
  a(B) = a_{\textrm{bg}} %
  \left(%
    1- \frac{\Delta B}{B-B_{0}}
  \right)%
\end{equation} %
where $B_{0}\approx 589.1$~G is the position of the Feshbach
resonance, $\Delta B \approx (1.4\pm 0.1)$~G is its width and
$\abg \approx 100\,\aBohr$ is the background scattering length, with
$\aBohr$ the Bohr radius. The magnetic field is directed along the
$z$-direction and determines the orientation of the dipoles. Using the
experimental procedure described in~\cite{Koch:08}, a condensate of
approximately $50,000$ atoms is formed in a crossed far-detuned
optical dipole trap at $\aevap\approx 85\,\aBohr$ above the Feshbach
resonance. We then shape the external confining potentials to obtain
the desired ratio of the trapping frequencies
$\lambda := \omega_z/\omega_y$ by adjusting the power in the crossed
dipole trap and superimposing (only for pancake-shaped traps
$\lambda>1$) an additional one-dimensional optical lattice along the
$z$-direction.  The superimposed lattice is formed by two far-detuned
beams (wavelength $\lambda_{\textrm{laser}}=1064$~nm) crossing under a
small angle ($\vartheta\approx 8^{\circ}$), as shown in
figure~\ref{fig:Exp_Setup}. This results in a relatively large spacing
of
$d_{\textrm{lat}}=\lambda_{\textrm{laser}}/[2\,\sin{(\vartheta/2)}]=(7.4\pm0.2)~\mu$m
between neighboring lattice sites.

We first adiabatically ramp the current in the offset coils linearly
in 8~ms to a scattering length $\ai$ close to the point where the
collapse occurs, and wait for 1~ms for the eddy currents (mainly due
to copper gaskets in our experimental chamber) to faint out. Thus, the
magnetic field $B(t)$ at the position of the atoms does not follow the
ramp instantaneously, but must be calculated according to the equation
\begin{equation} %
  \label{eq:Eddy_Currents} %
  B(t) = \tilde B(t) - \tau_B \dot B(t) %
\end{equation} %
where $\tilde B(t)$ is the magnetic field produced by the offset coils
(changed linearly during ramps) and $\tau_B \approx (0.57\pm 0.05)$~ms
is the $1/e$ lifetime of the eddy currents, measured by Zeeman
spectroscopy~\cite{Lahaye:08}. After 1~ms of waiting time at $\ai$, we
start a second ramp from $\ai$ to $\af<\acrit(\lambda)$, where
$\acrit(\lambda)$ is the critical scattering length for the given
trapping potential~\cite{Koch:08}, so that the collapse occurs.  We
hold the atoms in the trap for an additional time $\thold$ at $\af$
before releasing them and taking a time-of-flight image.  In order to
get the maximal absorption cross section, we split the time-of-flight
into two parts: a first one, lasting 4~ms, at the magnetic field
corresponding to $\af$ (in order not to disturb the dynamics) and a
second part, lasting again 4~ms, where the large magnetic field along
$z$ is replaced by a field of 11~G along the $x$-direction. We checked
that this procedure does not disturb the image.

The observed integrated density distribution is bimodal and consists
of a broad isotropic thermal cloud and an anisotropic
remnant BEC (see e.g. Fig~1(b) of~\cite{Lahaye:08}).  Because the size
of the thermal cloud as well as its atom number does not depend on
$\thold$, it is unlikely to contribute to the collapse dynamics. In
the following, we have subtracted it from the images to increase the
contrast.

\section{Collapse dynamics for different trap geometries}
The anisotropic character of the dipolar interaction (dipoles
side-by-side repel each other, while dipoles in a head-to-tail
configuration attract each other) has a strong effect on the stability
of a dipolar condensate: varying the geometry of the confining
harmonic trap from prolate to oblate (the symmetry axis being the one
along which the dipoles are aligned) stabilizes the condensate, as was
demonstrated experimentally in~\cite{Koch:08}. Our previous
experimental study of the collapse dynamics of a dipolar
condensate~\cite{Lahaye:08} was restricted to an almost spherical
trap; in the following we study the influence of the trap geometry on
the collapse dynamics.

\begin{figure}[t]
  \centering
   \includegraphics[scale=1]{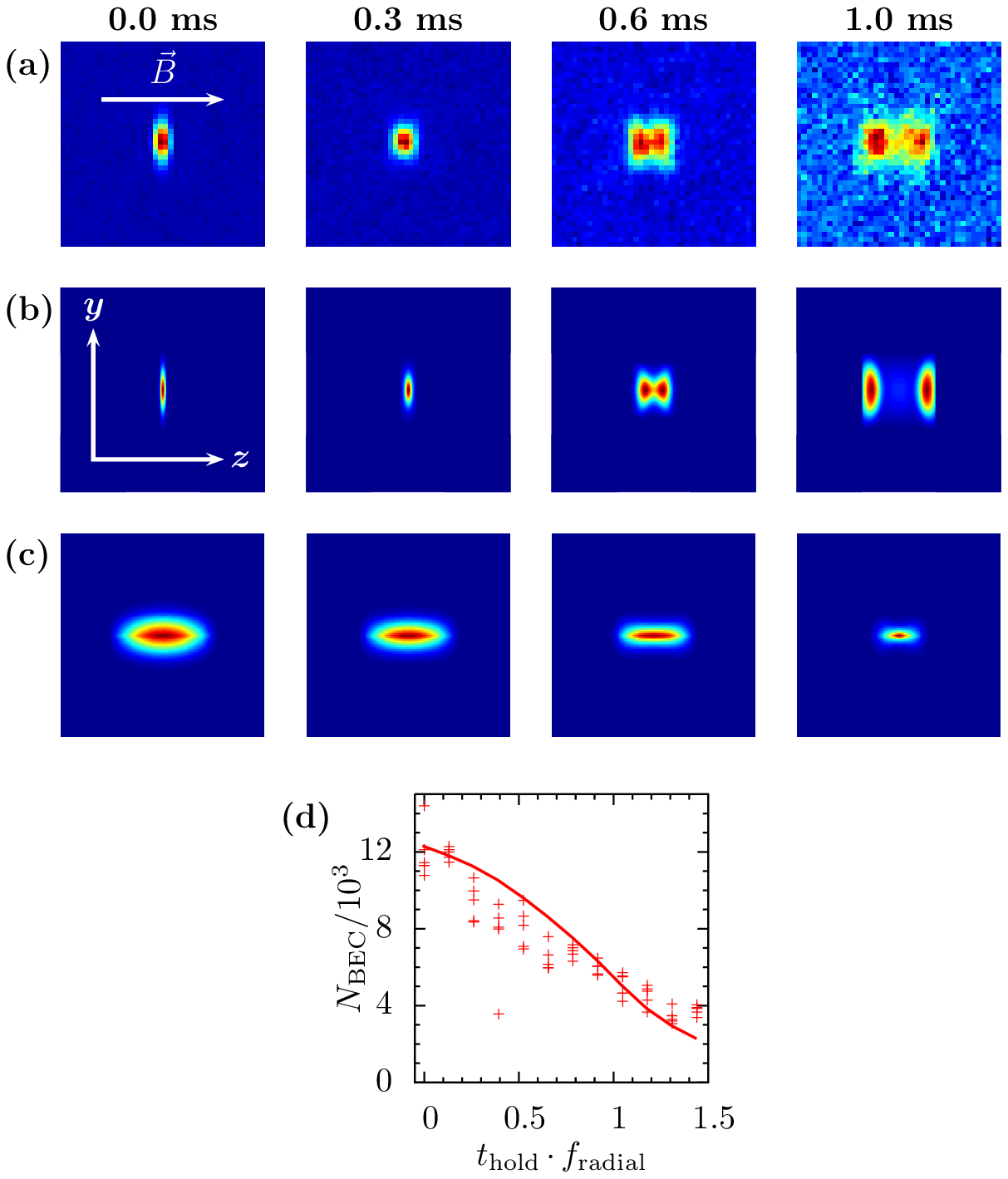}
   \caption{
     Comparison of the experimental absorption images %
     (a) with the simulations %
     (b) for a cigar-shaped trap with trapping frequencies
     $(f_x, f_y, f_z) = {(1312, 1311, 161)}$~Hz and trap ratio
     $\lambda\approx 0.12$.  Each image shows the averaged column
     density of five pictures after 8~ms of time-of-flight. The field
     of view is $250\times250$~$\mu$m$^2$. The final value of $\af$
     was $(8\pm3)\,\aBohr$ (also in the simulation). %
     (c) Simulated in-trap absorption images. The field of view is
     $(y,z)=(4.8, 25)\,\mu{\rm m}$.  %
     (d) The remnant condensate atom number for different holding
     times in units of the radial trap period
     $\tau=1/f_{\textrm{radial}}\simeq0.76$~ms. The solid line is the
     result of the numerical simulation (see text), without any
     adjustable parameter.  }
  \label{fig:not-collpased-state}
\end{figure}

\subsection{Prolate traps}
For purely contact interacting condensates the time scale which
  governs the ``usual dynamics'' is set by the largest trap
  frequency. In contrast, for dipolar condensates this time scale is
  given by the largest \emph{radial} trap frequency
  $\tau := 1/f_{\textrm{radial}}$, because the collapse is induced in
  this direction~\cite{Lahaye:08}.  %

Figure~\ref{fig:not-collpased-state} presents the collapse of a
dipolar condensate in a very elongated prolate trap
($\lambda\simeq0.12$).  Here we ramp from an initial scattering length
$\ai={(35\pm2)}\,{\aBohr}$ to $\af={(8\pm3)}\,{\aBohr}$, which lies
below the critical scattering length $\acrit\approx{12}\,{\aBohr}$. On
the time scale of the radial trap period $\tau \approx 0.76$~ms the
condensate only starts to split.  This is explicitly shown in
figure~\ref{fig:not-collpased-state}~(d): The atom number does not
drop ``abruptly'' to its final value, but instead it changes linearly
on the time scale $\tau$. Therefore, in the case of
figure~\ref{fig:not-collpased-state} we observe only a ``moderate''
collapse.

\begin{figure}[t]
  \centering
  \includegraphics[scale=1]{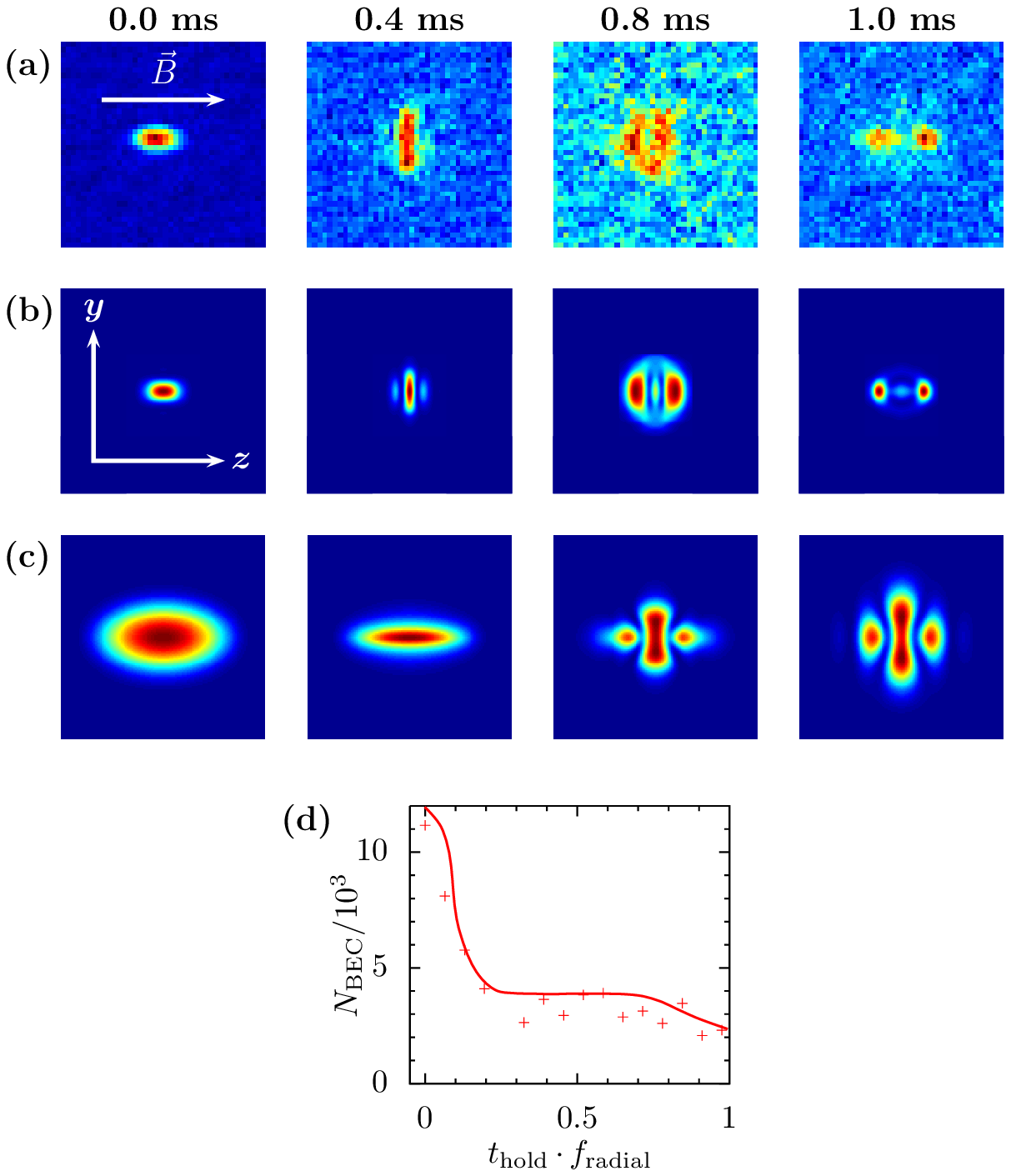}
  \caption{ (a, b) Collapse dynamics for different holding times
    $t_{\textrm{hold}}$ in a weakly cigar-shaped trap with trap
    frequencies %
    $(f_x, f_y, f_z) \approx {(650, 520, 400)}$~Hz, trap ratio
    $\lambda\approx0.7$ and radial trap period $\tau\sim1.7$~ms.  Each
    picture is a single absorption image after 8~ms of time-of-flight
    and has its own optical density scale. The field of view is
    $250\times250\;{\mu}$m$^2$. %
    (c) Simulated in-trap absorption images. The field of view is $6.9 \times 6.9\,\mu{\rm m}^{2}$. %
    (d) $N_{\rm BEC}$ versus $t_{\rm hold}$. The solid line is the
    simulation result.}
  \label{fig:5-50-0collapse_pics}
\end{figure}

Each panel in figure~\ref{fig:not-collpased-state}~(a) and (b) presents the
integrated column density $\int dx |\psi(\vec r, t)|^{2}$ of an
absorption image after 8~ms of free expansion. While the upper row
shows the experimental data, the lower rows is obtained from a
numerical simulation of the three dimensional Gross-Pitaevskii
equation %
\begin{equation*} %
    \label{eq:GPE} %
    \fl 
    i \hbar \frac{\partial}{\partial t} \psi(\vec r, t) = %
    \Bigg\{%
     -\frac{\hbar^{2}}{2m} \nabla^{2} %
      + V_{\textrm{trap}} %
      + \int U(\vec r - \vec r^{\,\prime}, t) %
      |\psi(\vec r^{\,\prime}, t)|^{2} {\rm d}^{3} r^{\prime} %
      - \frac{i\hbar L_{3}}{2} |\psi(\vec r, t)|^{4} %
    \Bigg\} %
    \; \psi(\vec r, t) %
\end{equation*} %
with the contact and dipolar interactions %
\begin{equation*} %
  U(\vec r,t) = %
  \frac{4\pi\hbar^{2} a(t)}{m} \; \delta(\vec r) %
  + \frac{\mu_{0} \mu^{2}}{4\pi} \; %
  \frac{1-3\cos^{2}{\theta}}{|\vec r|^{3}} %
\end{equation*} %
where $m$ is the atomic mass, $\mu_{0}$ the magnetic permeability of
vacuum, $\mu=6\mu_{\textrm{Bohr}}$ the magnetic moment of chromium,
and $\theta$ is the angle between $\vec r$ and the magnetic field
$\vec B \parallel \vec e_{z}$. The imaginary term describes the atom
losses. The simulations contain \emph{no} free parameter, as the
three-body coefficient
$L_{3}={2\times10^{-40}}\,{{\rm m}^{6}/{\rm s}}$ was estimated from
measurements.  Details about the simulations are described
in~\cite{Lahaye:08}.

Figure~\ref{fig:5-50-0collapse_pics} shows the collapse in a
cigar-shaped trap with trapping frequencies %
$(f_x, f_y, f_z) \approx {(650, 520, 400)}$~Hz,
corresponding to  $\lambda\simeq0.7$, and
radial trapping period  $\tau\simeq1.7$~ms.  We start with
$\Nbec = 13,500 \pm 1,500$ atoms before ramping the scattering length
from $\ai={(35 \pm2)}\,{\aBohr}$ to $\af={(8 \pm 3)}\,{\aBohr}$, which
lies ${4}\,{\aBohr}$ below the critical scattering length.  The
absorption images indicate three different stages: First, for
$\thold={0}$~ms, the condensate is strongly elongated along the
magnetic field direction, demonstrating strong dipolar
interactions~\cite{Lahaye:07}. Second, we observe a change of
ellipticity after ${0.4}\,{\rm ms}\approx{0.3}\,{\tau}$. This is a
consequence of the radial implosion and subsequent explosion (a
stable, cigar-shaped dipolar condensate does not invert its
ellipticity during the free expansion~\cite{Lahaye:07}). Third, we
observe a splitting of the condensate in axial direction as in
figure \ref{fig:not-collpased-state}, but now after
${0.8}\,{\rm ms}\approx{0.5}\,{\tau}$.  For longer holding times the
splitting becomes more prominent, but the dynamics is already
completed after half a trapping period.  This is shown in
figure~\ref{fig:5-50-0collapse_pics}~(d). The remnant atom number
$\Nbec(t)$ drops from $12,000$ to $4,000$ within this time scale. Note
that, for this figure (and only this one), the simulation results
agree better with the data when performed with $\af=2\,\aBohr$ (the
results shown in figure~\ref{fig:5-50-0collapse_pics} are for this
value of $\af$). This slight discrepancy is most probably due to a
slow drift of the magnetic field that occurred between the data
acquisition and the calibration of the scattering length (the
calibration procedure, described in details in~\cite{Lahaye:07},
requires the accumulation of a large quantity of data).

\subsection{Oblate trap}
\begin{figure}[t]
  \centering
  \includegraphics[scale=1]{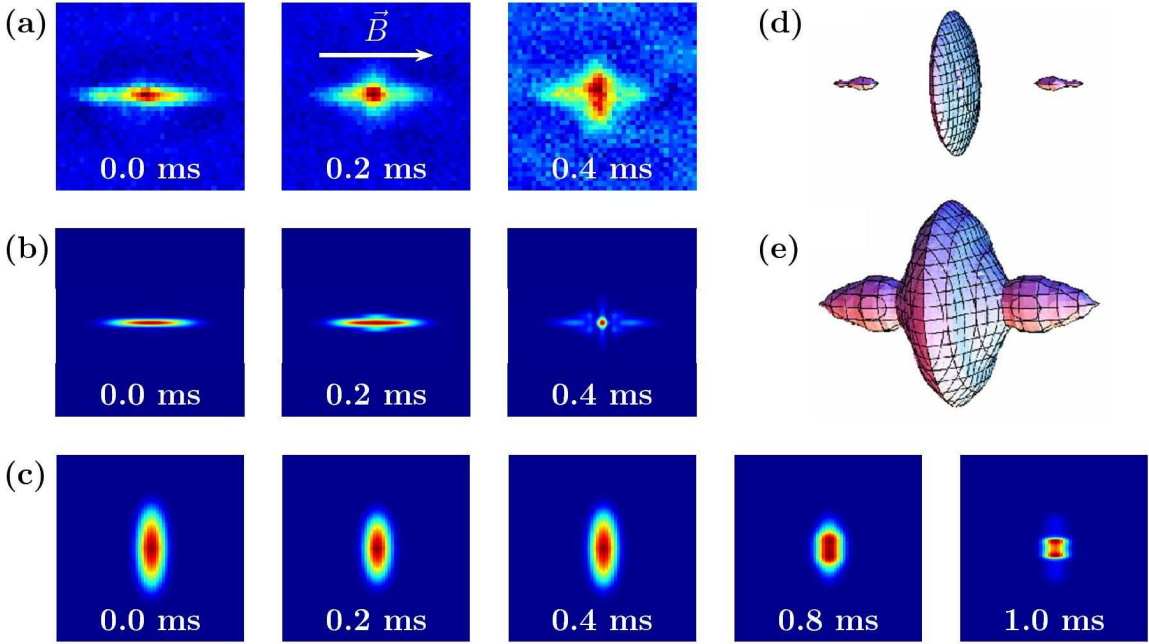}
  \caption{%
    (a, b) The collapse dynamics for different holding times $\thold$
    in a pancake-shaped trap, which corresponds to trap-frequencies
    $(f_x, f_y, f_z) \sim {(400, 400, 3400)}$~Hz, trap ratio
    $\lambda\approx8.5$ and radial period $\tau\simeq2.5$~ms.  The
    upper row present the average of five absorption images after 8~ms
    of free expansion.  The final ramp starts at
    $\ai={(30\pm2)}\,{\aBohr}$ and stops at
    $\af={(-13\pm2)}\,{\aBohr}$. The simulations give
    $\acrit={(-1.5 \pm 0.5)}\,{\aBohr}$. %
    (c) Simulated in-trap absorption images. The field of view is
    $(y,z)=(8.9, 4.5)\,\mu{\rm m}$.  %
    Isodensity surfaces for ``high'' %
    (d) and ``low'' %
    (e) densities of the $\thold=0.4$~ms image obtained from
    the Abel transformation. %
  }
  \label{fig:3-0-84collapse_pics}
\end{figure}

As discussed,  different trap
geometries are expected to result in different collapse dynamics. So
figures~\ref{fig:not-collpased-state} and
\ref{fig:5-50-0collapse_pics} have to be compared to the time
evolution of an almost spherical and a pancake-shaped trap.  While the
collapse in an almost spherical trap is published in~\cite{Lahaye:08},
we present the pancake-shaped trap in
figure~\ref{fig:3-0-84collapse_pics}.

The pancake-shaped trap is formed by superimposing  two additional
lattice beams onto a condensate, which was produced in the crossed
dipole trap, see figure~\ref{fig:Exp_Setup}. Depending on the
non-stabilized relative phase of the two lattice beams, it is in
principle possible that the condensate splits into two. Practically,
we have never observed interference fringes, even at expansion times
long enough, so that the fringe spacing
$h t_{\textrm{tof}}/(m d_{\textrm{lat}})$ was larger than our
${6}\,{\mu}$m resolution limit.  Switching off the trapping potential
immediately after finishing the magnetic field ramp results in a very
elongated condensate, in agreement with the simulations
(figure~\ref{fig:3-0-84collapse_pics}).

The cylindrical symmetry of the pancake-shaped trap allows us to
recover the three-dimensional density distribution $n(\rho,z)$ from
the two-dimensional absorption image
$n_{\textrm{abs}}(y,z)$. Figure~\ref{fig:3-0-84collapse_pics}~(d) and
(e) show isodensity surfaces for two different densities obtained from
the inverse Abel transformation~\cite{Bracewell:book} %
\begin{equation} %
  \label{eq:AbelTrafo} %
  n(\rho,z) = \frac{1}{2\pi} %
  \int_0^\infty dk \cdot k \cdot J_0(k \cdot \rho) %
  \int_{-\infty}^{\infty} d y \, %
  n_{\textrm{abs}}(y,z) \exp{(-i k \cdot y)} %
\end{equation} %
for $\thold={0.4}$~ms, where $J_0$ is the Bessel function of the first
kind.  As in the almost spherical
trap~\cite{Lahaye:08}, the collapsed cloud exhibits a \mbox{$d$-wave}
symmetry. While the isodensity surface for ``high'' densities contains
a central disc and two separated ``blobs'' along the symmetry axis,
the three parts merge for ``low'' densities.  Because the Abel
transformation is very sensitive to noise, we cannot extract in a reliable way the
kinetic energy released during the collapse from our images.

\section{Testing the coherence of the collapsed cloud\label{sec:Testing the coherence of the collapsed cloud}}
By analyzing the collapse of condensates confined in different trap
geometries, we showed that the survival density pattern exhibits two
parts: one part, which is well described by a thermal cloud and a
second part, which, due to its high optical density, we interpreted as
a remnant condensate~\cite{Lahaye:08}. In order to confirm this interpretation, we
checked the coherence of interfering condensates in the case of
pancake-shaped traps.

\begin{figure}[t]
  \centering
  \includegraphics[scale=1]{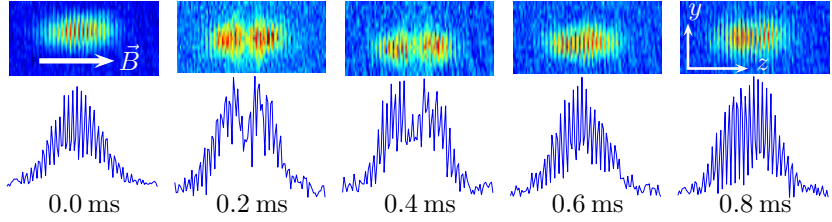}
  \caption{Interference pattern of three to five independent condensates for
    different holding times $\thold$ in the pancake-shaped trap,
    $(f_x, f_y, f_z) \sim {(400, 400, 3400)}$~Hz. The lower graphs show the
    $y$-integrated column densities $\sum_{y}\int dx |\psi|^{2}$
    (time-of-flight ${18}$~ms, field of view $110 \times 690\,{\mu}$m$^2$).
   }
  \label{fig:interferences}
\end{figure}

For that purpose, we produce several condensates by superimposing the
two beams for the optical lattice \emph{before} finishing the
evaporation to reach quantum degeneracy. Here the extension of the
cloud is still large enough, so that the atoms occupy three to five
adjacent lattice sites. The exact number of occupied sites changes
from shot to shot, depending on the relative phase of the two lattice
beams, which are not actively stabilized. After condensing, the BECs
on different lattice sites have random phases with respect to each
another. The single particle tunneling rate is vanishingly small
($<10^{-30}/$s), so that no phase coherence between adjacent sites is
built up on the time scale of the experiment.

It is known~\cite{Hadzibabic:04} that the interference fringes are not
washed out if few independent condensates interfere instead of only
two. Taking the same trap geometry as in
figure~\ref{fig:3-0-84collapse_pics}, but extending the time-of-flight
to 18~ms, we obtain the interference patterns shown in
figure~\ref{fig:interferences}.  As expected, the absolute position of
the interference fringes changes from shot to shot, but they can be
clearly seen on each image.

While for holding times shorter than $0.2$~ms or longer than
  $0.5$~ms interference fringes with a high contrast are
  visible, we observe no interference fringes at the center of the two
  clouds for $0.2\textrm{~ms}\le \thold \le 0.5$~ms. A possible
  interpretation is as follows: For $\thold=0$~ms \emph{no} collapse
  occurs and the observed fringes are similar to those of two point
  sources. For $\thold=0.2$~ms and $0.4$~ms the condensates do
  collapse, but this happens during the time-of-flight and
  \emph{after} the clouds overlapped: e.g.\ for $\thold=0.4$~ms the
  condensates start to overlap for $t_{\textrm{tof}} = 0.4$~ms, but
  the collapse happens at $t_{\textrm{tof}} = 0.8$~ms. Probably this
  induces a complicated phase distribution in each of the condensate
  and integration over the line of sight washes out the interference
  fringes. On the other hand, if the collapse happens in-trap (e.g.\
  for $\thold = 0.8$~ms) the fringes are formed by the remnant
  condensates. Again we recover the fringes as if the atoms would
  belong to two coherent point sources.

\section{Conclusions}
We have investigated theoretically and experimentally the collapse
dynamics of dipolar condensates in prolate and oblate harmonic
trapping potentials. As expected, the collapse dynamics depends on the
trap geometry, although the qualitative behavior is similar for
different traps. The simulations containing no adjustable parameter
reproduce the experimental results well. By simultaneously inducing a
collapse in several condensates and let them interfere, we showed that
the collapsed cloud contains a coherent remnant condensate.

A clear direction of further studies is to use the interferometric
technique of
section~\ref{sec:Testing the coherence of the collapsed cloud} to
obtain experimental evidence for the vortex rings predicted
in~\cite{Lahaye:08}.  The contrast of the interference fringes is not
high enough to support their evidence yet. An other interesting
extension is the study of two-dimensional
solitons~\cite{Pedri:05,Tikhonenkov:08}, which are expected to appear
just above the instability threshold.

\ack
We acknowledge support by the German Research Foundation (SFB/TRR 21),
the Landesstiftung Baden-W{\"u}rttemberg, and the EU (Marie-Curie
Grant No. MEIF-CT-2006-038959 to T.\ L.). H. S., Y. K., and
M. U. acknowledge support by the Ministry of Education, Culture,
Sports, Science and Technology of Japan (Grants-in-Aid for Scientific
Research No. 17071005 and No. 20540388, the Global COE Program "the
Physical Sciences Frontier") and by the Matsuo Foundation.


\end{document}